\documentclass[preprint,double-spaced,showpacs,preprintnumbers,
amsmath,amssymb,overcite]{revtex4}
\usepackage{graphicx}% Include figure files
\usepackage{dcolumn}% Align table columns on decimal point
\usepackage{bm}% bold math

\begin{document}

\title{Photo- and electro-production of medium mass
       $\Lambda$-hypernuclei}

\author{P.~Byd\v{z}ovsk\'y$^1$, M.~Sotona$^1$, T.~Motoba$^2$,
        K.~Itonaga$^3$, K.~Ogawa$^4$, and O.~Hashimoto$^5$}
\affiliation{
    $^1$Nuclear Physics Institute, 250 68 \v{R}e\v{z}
        near Prague, Czech Republic\\
    $^2$Laboratory of Physics, Osaka Electro-Communication
        University,  Neyagawa 572-8530, Japan\\
    $^3$Laboratory of Physics, Faculty of Medicine,
        University of Miyazaki,
        Miyazaki 889-1692, Japan\\
    $^4$Department of Physics, Chiba University, Chiba 263-8522,
        Japan\\
    $^5$Department of Physics, Tohoku University,
        Sendai 980-8578, Japan}

\date{\today}
\begin{abstract}
The characteristic and selective nature of the electro-magnetic
production of $\Lambda$-hypernuclei in exciting states is
demonstrated assuming the medium-mass targets $^{28}$Si,
$^{40}$Ca, and $^{52}$Cr. In the analysis, formalism of DWIA is
used adopting the Saclay-Lyon, Kaon-MAID, Adelseck-Saghai, and
Williams-Ji-Cotanch models for the elementary production process
and various nuclear and hypernuclear wave functions. The
elementary amplitudes are discussed in detail presenting their
basic properties and comparison with data. The unique features of
the electro-magnetic production of $\Lambda$-hypernuclei shown on
the examples are the selective excitation of unnatural parity
highest-spin states (natural parity ones for the $LS$-closed
targets) and a possibility to investigate the $\Lambda$
single-particle energies including a spin-orbit splitting using
variety of medium-mass targets.

\end{abstract}
 \maketitle

\section{Introduction}

The hypernucleus is a strongly-interacting multibaryonic system
with hyperon(s) and its lifetime is as long as the typical order
of the $\Lambda$ weak-interaction decay ($\tau \approx 10^{-10} $
sec). Thus a variety of hypernuclear phenomena provide us with new
knowledge of baryon-baryon interactions and novel behaviors of
many-body structures. Among these systems the
$\Lambda$-hypernuclei have been relatively well produced mainly
through the $({\rm K}^-,\pi^-)$ and $(\pi^+,{\rm K}^+)$ reactions.
However, as these experiments have more or less limitation due to
their own kinematics, more variety of hypernuclear production
processes should be challenged in order to get information on
detailed properties of hyperon-nucleon interactions. In this paper
we like to demonstrate a new and wide possibility of the photo-
and electro-production of hypernuclei, by taking the actual
experimental facilities into account, and will show a novel aspect
of the reaction spectroscopy.

Historically, the stopped K$^-$ absorption reaction was utilized
to produce $\Lambda$-hypernuclei of which decays were identified
in bubble chamber and nuclear emulsion. Next, the in-flight
reaction ${\rm K}^-+ {\rm n} \rightarrow \pi^-+ \Lambda$ with the
kaon beam momenta $p_{\rm K}=400 - 800 ~{\rm MeV}/c$ had a unique
role of getting particular hypernuclear excited states taking an
advantage of the recoilless condition ($\Delta L=0$) with very
weak spin-flip transition. Then a new stage of hypernuclear
production was realized by employing the $\pi^+ + {\rm n}
\rightarrow {\rm K}^+ + \Lambda$ reaction mostly at $p_\pi \approx
1.05 ~{\rm GeV}/c$. This $(\pi^+, {\rm K}^+)$ reaction on nuclear
targets played a great role of producing a series of all $\Lambda$
single-particle states inside the nucleus, since the process with
recoil momentum $q \approx 350 ~{\rm MeV}/c$ preferentially
excites hypernuclear high-spin stretched configurations with
natural parity mainly due to the spin-non-flip component of the
elementary amplitudes. More details  can be found in the review
papers~\cite{HaTa06,BMZ90,Springer}.

In general, hypernuclear states consist of multiplets which are
based on the nuclear core angular momentum $J_{\rm c}$ coupled
with that of a hyperon $j_{\Lambda}$, so that low-lying doublets
with a $s$-state $\Lambda$ have spin partners: $J_{\rm H}=J_{\rm
c} \pm 1/2$. The $({\rm K}^-,\pi^-)$ reaction with spin-non-flip
nature leads to excitation of one of the doublet member with
natural parity. Although the $(\pi^+,{\rm K}^+)$ spin-flip
component is sizable enough to produce appreciable polarization in
the final hypernuclear states~\cite{Bando}, actually the favored
states with large cross section are solely restricted to the
natural parity states and their unnatural parity partners get very
small cross section~\cite{Bando,Ito94}. In addition the typical
energy resolution achieved in the $(\pi^+,{\rm K}^+)$ reaction
spectroscopy is 1.45 MeV at best~\cite{ref4x}, which is much
larger than the characteristic spin-doublet splittings of the
order of a few tens keV. Therefore, only the energy of natural
parity member of the doublets can be accessible within the
restriction of such energy resolution. The coincidence
$\gamma$-ray measurement proved to be a nice tool to know the
detailed hypernuclear level energies with a remarkably good
resolution of several keV~\cite{HaTa06}.

In contrast to the production processes mentioned above, which
convert a neutron into $\Lambda$, here we focus our attention
on K$^+$ photo- and electro-production process. These reactions
convert a proton in nuclear targets into $\Lambda$, so that new
variety of hypernuclei can be produced. Indeed much larger amount
of spectroscopic data for other hypernuclei with various excited
states are quite necessary to draw some clear conclusion about
the nature of hyperon-nucleon interaction for which the elementary
scattering experiments are not available. Over 10 years ago, three
of the present authors predicted the photo-production cross
sections for $^7$Li, $^{10}$B, $^{12}$C, and $^{16}$O
targets~\cite{SF,MotobaPTP117}. Among others the $^{12}$C case
has been remarkably well confirmed by the first experiment on
the nuclear target at the Thomas Jefferson National
Laboratory (JLab)~\cite{C12eeK}.

These factors explain the reasons why recently the
photo- and electro-production of both hyperons and hypernuclei
has attracted much attention in strangeness nuclear physics.
In addition the following facts encourage us to extend the
theoretical estimates to wider possibilities of precise
reaction spectroscopy:

\begin{itemize}
\item The quality of CEBAF beam (high intensity and energy
resolution) at JLab makes it possible to identify various
hypernuclear energy levels with appreciable cross sections
together with a resolution of about 400
keV~\cite{HaTa06,Hungerford117}.

\item The photo- and electro-production reactions, ($\gamma, {\rm
K}^+$) and (e,e'K$^+$), are characterized  by the large momentum
transfer ($q \ge$ 350 MeV/$c$) and the strong spin-flip terms.
This means especially that photons (real or virtual) have unique
characteristics of exciting unnatural parity high-spin
hypernuclear states preferentially including states with
deeply-bound $\Lambda$ hyperon~\cite{Sotona}.

\item In contrast to $({\rm K}^-, \pi^-)$ or ( $\pi^+, {\rm K}^+$)
reactions, the electro-magnetic production of K$^+ \Lambda$ pair
on the proton makes it possible to study some ``proton-deficient''
hypernuclear species, such as $^7_\Lambda$He, $^9_\Lambda$Li
and $^{12}_{\,\Lambda}$B, which are not available otherwise.
Such ability opens a new possibility of producing
neutron-rich hypernuclei with large excess of neutral baryons.
The same hypernuclei may be produced in $({\rm K}^-,\pi^0)$
reaction~\cite{BNL}, but due to the weak spin-flip interaction
in the latter the two reactions will afford complementary
information on hypernuclear spectroscopy.

\end{itemize}

The possibility to use the electro-production of strange particles
as a tool for study of hypernuclei was first mentioned in a
pioneering work by Fetisov {\it et al.}~\cite{fetisov}. In 1980's,
early works on photo- and electro-production of hypernuclei have
been done based on simplified models~
\cite{bernstein,cotanch1,rosenthal1,bennhold1}. The production
cross section and polarization of produced hypernuclei in
photo-production process was examined in
Refs.~\cite{MotobaPTP117,sotona1,richter} for schematic as well as
for very realistic shell-model wave functions and with full
inclusion of distortion effects. Electro-production of $0p$-shell
hypernuclei was investigated carefully in Refs.~\cite{SF,adam}
with close link to the JLab experimental
program~\cite{CEBAF1,CEBAF2,CEBAF3}.

The aim of this paper is, first, to demonstrate the characteristic
and selective nature of the photo-production reaction in exciting
hypernuclear states. For this purpose we adopt a typical
medium-mass target of $^{28}$Si and discuss the general and novel
rules based on the properties of the elementary amplitudes. We
also make clear the effect of kaon distortion and the different
hypernuclear wave functions. Second, we examine the properties of
the recent models of elementary kaon electro-magnetic production.
Third, by choosing medium-heavy nuclear targets such as $^{28}{\rm
Si}$, $^{40}{\rm Ca}$ and $^{52}{\rm Cr}$, we present theoretical
predictions of excitation spectra corresponding to the
hypernuclear programs at the JLab facility~\cite{CEBAF3}.

This paper is organized as follows: Section 2 briefly describes
the minimum framework of calculation of the hypernuclear
production cross section. In order to demonstrate the successful
prediction done 10 years ago, the updated excitation spectrum for
$^{12}$C(e,e'K$^+$)$^{12}_{\,\Lambda}$B is shown. In section 3 the
basic properties of modern elementary amplitudes of kaon
electro-magnetic production are discussed. Section 4 summarizes
the excitation spectra predicted for the $(\gamma, {\rm K}^+)$
reaction on $^{28}{\rm Si}$. In order to show the novel and
general selectivity found for the $(\gamma, {\rm K}^+)$ reaction,
the simple configuration estimates are presented in subsection
4.1, while sophisticated wave functions are employed in subsection
4.2 for the realistic predictions. In section 5, the calculation
is further applied to heavier hypernuclear production by choosing
the $^{40}{\rm Ca}$ and $^{52}{\rm Cr}$ targets. Conclusions and
summary are given in section 6.

%
% Section 2
%
\section{Formalism for hypernuclear production}

\begin{figure}[htb]
\begin{center}
  \includegraphics[height=8.6cm,angle=270]{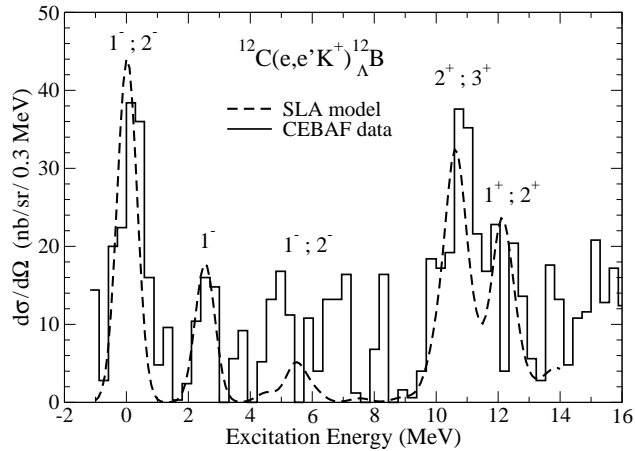}
\end{center}
\caption{Comparison of the experimental excitation spectrum for
the $^{12}$C(e,e'K$^+$)$^{12}_{\,\Lambda}$B reaction~\cite{C12eeK}
with the theoretical calculation done for real photons (see text)
at $E_\gamma = 1.42$\,GeV and $\theta_{\rm K}^{\rm L}= 2^\circ$
using the Saclay-Lyon A model. The comparison illustrates a
predictive power of calculations performed in the DWIA framework.}
\label{figure0}
\end{figure}

The cross section for the electro-magnetic production of
$^{12}_{\,\Lambda}$B hypernucleus was predicted already many years
ago~\cite{SF,MotobaPTP117} but until recently a good quality data~
\cite{C12eeK} allows for a comparison with experiment which
stimulates further development of the models. In
Fig.~\ref{figure0} we demonstrate that a very good agreement was
achieved for the $^{12}$C(e,e'K$^+$)$^{12}_{\,\Lambda}$B reaction~
\cite{C12eeK}. The calculation presented in Fig.~\ref{figure0} was
performed for the real photon whereas the measurement done at JLab
is for the virtual photons. However, kinematics of this experiment
is very close to the photo-production case because the photon
momentum squared ($q^2\equiv -Q^2$) is $-2.6\times
10^{-6}$\,(GeV/$c$)$^2$ at the electron scattering angle
0.1$^\circ$. Thus the photo-production cross section is
practically identical to the one for the electro-production
process. The successful first experiment has opened a variety of
hypernuclear production by electron
beams~\cite{CEBAF1,CEBAF2,CEBAF3}, and accordingly it is
challenging to make theoretical predictions for wide range of
hypernuclear productions together with improved treatment.
%
% Figure 1
%

\begin{figure}[htb]
\begin{center}
  \includegraphics[width=8.6cm]{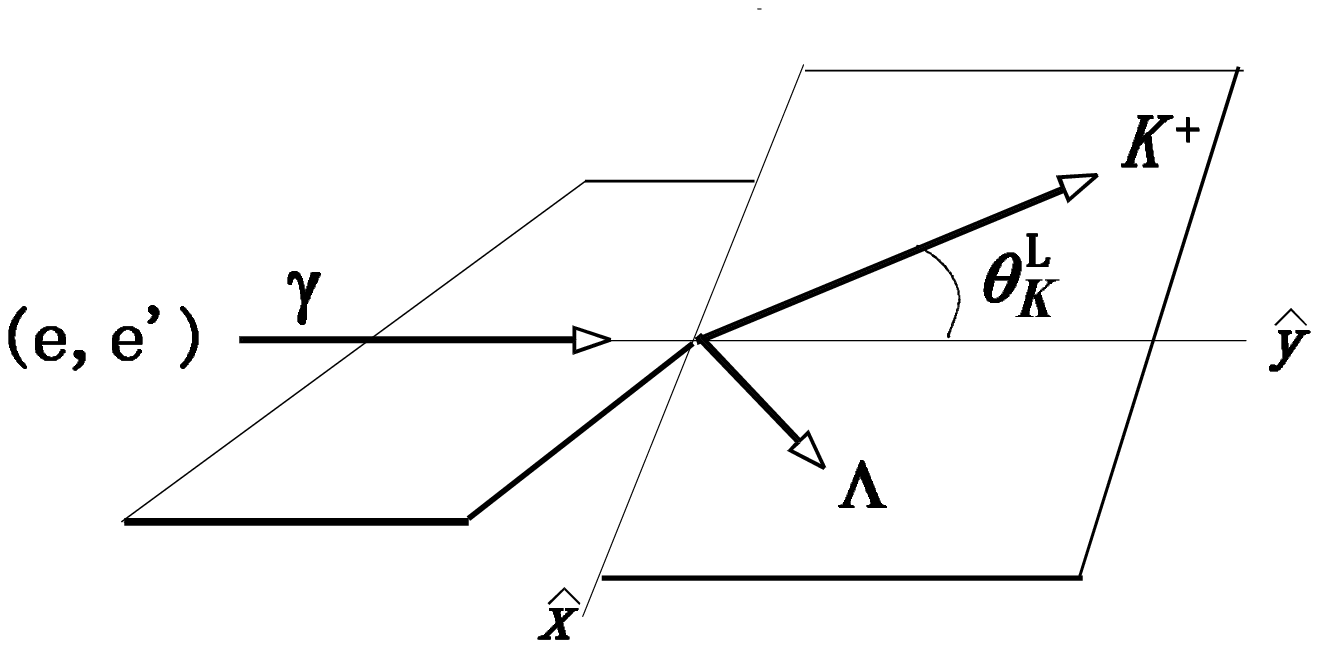}
\end{center}
\vspace*{-5mm}
\caption{Hyperon electro-magnetic-production kinematics in
the laboratory frame. Real or virtual photons are assumed.}
\label{figure1}
\end{figure}

The production cross section and polarization of hypernuclear
states are evaluated by choosing the coordinate frame \{$S_2$\}
defined as follows (see Fig.~\ref{figure1})
\begin{equation}
   \mbox{$\hat{\bf z}$} = \mbox{{$\hat{\bf n}$}} =
    \frac {{\mbox{${\bf p}_\gamma$}}\times
        {\mbox{${\bf p}_{\rm K}$}}}
  {|{\mbox{${\bf p}_\gamma$}}\times {\mbox{${\bf p}_{\rm
  K}$}}|}\,,\ \
 \ \mbox{$\hat{\bf y}$} = \mbox{{$\hat{\bf p}_\gamma$}}\,,\ \
 \ \mbox{$\hat{\bf x}$} =
  \mbox{$\hat{\bf y}$} \times  \mbox{{$\hat{\bf z}$}}\ .
\end{equation}
%
% Figure 2
%

For the real photon, the hypernuclear production cross section
is expressed in DWIA as
\begin{equation}
 \frac {d\sigma}{d\Omega}(\theta^{\rm L}_{\rm K}) =
  \frac {s\, p^2_{\rm K} E_{\rm K} E_{\rm H} }
        {p_{\rm K} (E_{\rm H} +E_{\rm K}) - E_\gamma E_{\rm K}\,
{\rm cos} \theta^{\rm L}_{\rm K}}
       {\displaystyle\;{\sum_{M_f} R(fi; M_f)}}\ ,
\end{equation}
where $p$'s and $E$'s are the momenta and energies in the $A$-body
laboratory frame, and $s$ is the square of the sum of the energies
of $\gamma$ and proton in their center of mass system. $R(fi;
M_f)$ is a transition strength defined as
\begin{equation}
  R(fi; M_f) = \frac {1} {[J_i]}
   {\displaystyle{\sum_{M_i} | \langle J_f M_f T_f \tau_f |{\cal O} |
    J_i M_i T_i \tau_i \rangle |^2 }}\ ,
\end{equation}
where the magnetic subspace quantum number $M_f$ is kept explicitly
and the convention $[a] = 2a + 1$ is used.

The transition operator ${\cal O}$ for the $(\gamma,{\rm K}^+)$
reaction is written as
\begin{equation}
  {\cal O} = \int\! d^3{\bf r}\, \chi_{\rm K}^{(-)*}({\bf p},
  \xi {\bf r})\,\chi_{\gamma}^{(+)}({\bf k},
  {\bf r})
  \sum_{\nu=1}^A V_-^{(\nu)}\delta
  ({\bf r}-\eta {\bf}r_{\nu})
  \langle{\bf k}-{\bf p},{\bf p}|\,t\,|
  {\bf k},0\rangle_{\rm lab}\ ,
\end{equation}
where $\xi=M_{\rm A}/M_{\rm H}$ and $\eta=(M_{\rm H}-M_{\rm
A})/M_{\rm A}$ are introduced to take the recoil effect into
account. $V_-^{(\nu)}$ represents a V-spin lowering operator which
converts a proton into a $\Lambda$ hyperon and is zero for
neutron.  The $t$-matrix in the 2-body lab system can be given by
using the baryon spin operators as
\begin{equation}
%{\cal M}
{\mathcal{M}}\equiv \langle{\bf k-p,p}|\,t\,|{\bf k},0\rangle_{\rm
lab}= \epsilon_0(f_0+g_0\sigma_0)+ \epsilon_x(g_{1}\sigma_1
+g_{-1}\sigma_{-1})\ .
\end{equation}
Here $\epsilon_0$ and $\epsilon_x$ denote the photon polarization
in the $z$ and $x$ directions, respectively, and the coefficients
of $g_0$, $g_{1}$, and $g_{-1}$ are defined in Ref.~\cite{MotobaPTP117}.

The actual evaluation of $R(fi; M_f)$ is carried out in the
coordinate frame \{$S_1$\} defined by
\begin{equation}
  \mbox{$\hat{\bf z}$} = \mbox{$\hat{{\bf p}}_\gamma$},
 \ \mbox{$\hat{\bf y}$} = \mbox{{$\hat{\bf n}$}},
 \ \mbox{$\hat{\bf x}$} = \mbox{{$\hat{\bf n}$}}
        \times \mbox{{$\hat{\bf z}$}}\ ,
\end{equation}
which is more suitable for calculations of incident and
outgoing distorted waves.
The coordinate frame $\{ S_2 \}$ is obtained from $\{ S_1 \}$ by
the rotation ${\cal R}(\frac {\pi}{2}, \frac {\pi}{2},
\frac {\pi}{2} )$. The $R(fi; M_f)$ is, then, expressed in terms of
the ``reduced effective number" $\rho(fi; M_f)$ as
\begin{eqnarray}
  R(fi; M_f) &=& |f_0|^2 \rho^{ff}(fi; M_f)
     + |g_0|^2 \rho^{gg}(fi; M_f)
     + 2{\rm Re}[ f_0g_0^* \rho^{fg}(fi; M_f) ] \\
     & + &  |g_1|^2 \rho^{gg+}(fi; M_f)
     + \ |g_{-1}|^2 \rho^{gg-}(fi; M_f)
     + \ 2{\rm Re}[ g_1 g_{-1} \rho^{gg+-}(fi; M_f) ]\ . \nonumber
\end{eqnarray}

The expressions of $\rho^{ff}$, $\rho^{gg}$ and $\rho^{fg}$ are
formally the same as those in the $(\pi^+, {\rm K}^+)$ reaction,
so that one may refer to the Appendix of Ref.~\cite{Bando}. The
other quantities $\rho^{gg+}$, $\rho^{gg-}$, and $\rho^{gg+-}$ are
essentially new ones which appear in the $(\gamma, {\rm K}^+)$
reaction for the first time. Their expressions are listed in the
Appendix.

\section{Basic properties of the elementary kaon photo-production
  amplitudes}

\begin{figure}[htb]
\begin{center}
  \includegraphics[height=8.6cm,angle=270]{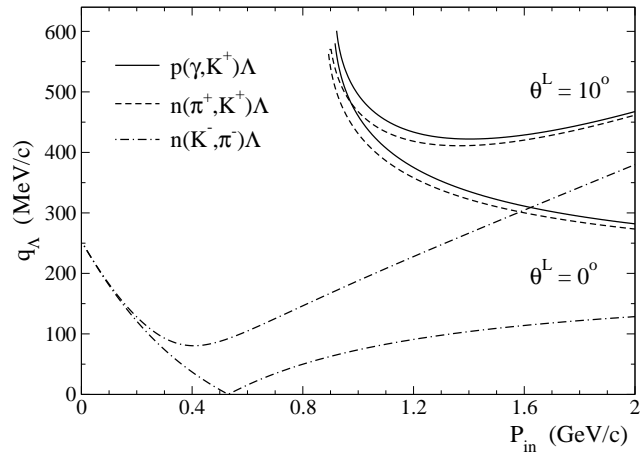}
\end{center}
\vspace*{-5mm}
\caption{ Momentum transfer to the $\Lambda$ hyperon is plotted
as a function of projectile laboratory momentum $P_{\rm in}$.
The two curves for each reaction correspond to the two values
of the kaon (pion) angle $\theta^{\rm L}$= 0$^\circ$ (lower curves)
and $\theta^{\rm L}$= 10$^\circ$ (upper curves).}
\label{figure2}
\end{figure}

In hypernuclear production a magnitude of the momentum transfer to
the $\Lambda$ hyperon plays an important role in exciting a series
of hypernuclear highest-spin states. Before discussing properties
of typical elementary amplitudes for kaon photo-production, we
compare in Fig.~\ref{figure2} the momentum transfer $q_{\Lambda}$
for ${\rm p}(\gamma,{\rm K}^+)\Lambda$, ${\rm n}(\pi^+,{\rm
K}^+)\Lambda$, and ${\rm n}({\rm K}^-,\pi^-)\Lambda$ reactions,
the last two having been often performed until present. In the
figure the two curves, which correspond to the kaon (pion) angle
$\theta^{\rm L} = 0^\circ$ (lower one) and $\theta^{\rm L} =
10^\circ$ (upper one), are shown for each reaction drawn as a
function of the incident momentum $P_{\rm in}$. In the following
hypernuclear calculations we choose $P_{\gamma}=1.3$ GeV/$c$,
which is close to the energy selected in several proposals of
experiments at JLab. The momentum transfer $q_{\Lambda}$, for the
photo-production, then amounts to 353 MeV/$c$ at $\theta^{\rm
L}$=0$^\circ$ and 425 MeV/$c$ at $\theta^{\rm L}$=10$^\circ$.
These values are well comparable to those for the ${\rm
n}(\pi^+,{\rm K}^+)\Lambda$ reaction at $P_{\pi}=1.05$ GeV/$c$:
406 ($\theta^{\rm L}$=0$^\circ$) and 447 ($\theta^{\rm
L}$=10$^\circ$) MeV/$c$. On the other hand, one may refer to
smaller momentum transfer involved in the ${\rm n}({\rm
K}^-,\pi^-)\Lambda$ reaction, $q_{\Lambda}$ = 50 and 147 MeV/$c$
at $P_{\rm K}= 0.8$ GeV/$c$ and $q_{\Lambda}$ = 109 and 285
MeV/$c$ at $P_{\rm K}= 1.5$ GeV/$c$.
%
% Figure 3
%

Numerous theoretical attempts have been made to describe the
elementary $\gamma {\rm p}\to \Lambda {\rm K}^+$ process. In the
kinematical region assumed here, $E^{\rm L}_{\gamma}=0.91-2$ GeV,
the isobaric models~
\cite{MDLS,AW88,AS90,WC92,SL96,SLA98,KMAID,Janxx} based upon the
Feynman diagram technique are of particular interest. In these
models the amplitude is derived from an effective hadronic
Lagrangian in the tree level approximation. It gains contributions
from the extended Born diagrams, where the proton, lambda,
$\Sigma^0$, and kaon are exchanged in the intermediate state, and
diagrams which account for exchanges of moderate mass (less than 2
GeV) nucleon, hyperon, and kaon resonances. Unfortunately, due to
absence of a dominant baryon resonance in the process~
\cite{Tho66}, on the contrary to the pion and eta production, many
of exchanged resonances have to be {\it a priori} assumed~
\cite{SL96} introducing a rather large number of free parameters
in calculations, the appropriate coupling constants. The free
parameters are determined by fitting the cross sections and
polarizations to the experimental data which, however, provides a
copious number of possible sets of parameters~\cite{AS90,SL96}.
This large number of models which describe the data equally well
can be reduced implementing duality hypothesis, crossing symmetry,
and SU(3) symmetry constraints.

According to duality principle most of the nucleon resonances
exchanged in the s-channel, especially those with a high spin, can
be mimic by the lowest-mass kaon poles K$^*$ and K$_1$ in the
t-channel~\cite{WC92,SL96}. The crossing symmetry constraint
requires in order that a realistic model for the $\gamma {\rm p}
\to \Lambda {\rm K}^+$ process yields simultaneously a reasonable
description of the radiative capture of K$^-$ on the proton with
$\Lambda$ in the final state~\cite{WC92,SL96}. The flavor SU(3)
symmetry allows to relate the main coupling constants $g_{\rm
K\Lambda N}$ and $g_{\rm K\Sigma N}$ to the better established
one,  $g_{\pi\rm NN}$. For the 20\% breaking of the symmetry the
following limits can be obtained for them: $-4.4 \leq g_{\rm
K\Lambda N}/\sqrt{4\pi} \leq -3.0$ and $0.8 \leq g_{\rm K\Sigma
N}/\sqrt{4\pi} \leq 1.3$~\cite{AS90,SL96}. Analysis of data
performed under different
assumptions~\cite{MDLS,AW88,AS90,WC92,SL96,SLA98} showed that a
moderate number of resonances is sufficient to get a reasonable
agreement with the experimental data.

In the isobaric models discussed above the baryons are assumed as
a point-like particles in the strong interaction vertices which is
forced by the gauge invariance principle. Recently, however,
baryon form factors were successfully introduced in the model in a
gauge-invariant way~\cite{HBMF98,WD}. Accounting for the composite
structure of the baryons proves to be important in suppression of
the  contribution of the Born terms~\cite{Janxx}.

More elementary approach to study of the reaction mechanism of
$\gamma {\rm p}\to \Lambda {\rm K}^+$ was performed in terms of quark
degrees of freedom in Refs.~\cite{ZpL95,LLP95}. These models being
in a closer connection with QCD than those based on the baryon
degrees of freedom, need a smaller number of parameters to describe
the data. Moreover, the quark models assume explicitly an extended
structure of the baryons which was found to be important for a
reasonable description of the photo-production data~\cite{LLP95}.
Other approaches based on the Regge trajectory formalism~\cite{Rgg97}
and chiral perturbation theory~\cite{Chpt} are not suited for using
here because they are applicable to photon energies larger than
3-4 GeV and to the threshold region, respectively.

In hypernuclear calculations performed here we adopt four isobaric
models denoted hereafter as  Kaon-MAID (KMAID)~\cite{KMAID},
Adelseck-Saghai (AS1)~\cite{AS90}, Williams-Ji-Cotanch
(C4)~\cite{WC92}, and Saclay-Lyon A (SLA)~\cite{SLA98} which
characterize nowadays understanding of the $\gamma {\rm p} \to \Lambda
{\rm K}^+$ process for $E_{\gamma}^{\rm L} \leq 1.5$ GeV. The models
KMAID, C4, and SLA are extended to the higher energies,
$E_{\gamma}^{\rm L} \leq 2.2$ GeV, and to the electro-production
process. They also aimed at description of the channels with
$\Sigma^0$ and $\Lambda(1405)$ (C4) or $\Sigma^+$ (KMAID, SLA) in
the final state. The models C4 and SLA describe also data on the
radiative capture of K$^-$ on the nucleon.

A common feature of the four models, KMAID, AS1, C4, and SLA, is
that besides the extended Born diagrams they include also the kaon
resonant ones, K$^*$(890) and K$_1$(1270). As shown in
Ref.~\cite{WC92} these t-channel resonant terms in combination
with s- and u-channel exchanges  improve an agreement with the
data in the intermediate energy region. The models differ in a
particular choice of nucleon and hyperon resonances. The  models
AS1, and C4 assume only the spin 1/2 baryon exchanges with
different masses, whereas in the more elaborate models, KMAID and
SLA, the spin 3/2 nucleon resonance N(1720) was added. Moreover,
to account for the resonant structure seen in the SAPHIR
data~\cite{SPH98} and confirmed in the latest
measurements~\cite{SPH03,CLAS} the KMAID model includes a
resonance $D_{13}$(1895) which was predicted by the quark
model~\cite{CR94} but which was not observed yet. The higher-spin
resonances were omitted in the C4 assuming the duality hypothesis
whereas in the Saclay-Lyon model they were introduced to improve
an agreement with the data at higher energies~\cite{SL96}. The
model SLA is a simplified version of the full Saclay-Lyon
model~\cite{SL96} in which the nucleon resonance with spin 5/2
appears in addition. Since predictions of the both models are very
similar for the cross sections and polarizations in the kaon
photo-production~\cite{SLA98} we have chosen the simpler version
SLA here.  The only KMAID model includes the baryon form factors
realized in the prescription by Haberzettl {\it et
al.}~\cite{HBMF98} which improves its predictions for higher
energies.

The electro-magnetic and strong coupling constants were fixed by
fitting to various sets of the experimental data which defines a
range of validity of the models. The model AS1 confines itself to
the kaon photo-production data for $E_{\gamma}^{\rm L} <$ 1.4 GeV,
violates the crossing principle by over-predicting the branching
ratio of the radiative capture~\cite{AS90} but it fulfills the
SU(3) symmetry limits for the two main coupling constants.
Parameters of the model C4 were fitted to the data on the kaon
photo- and electro-production and the radiative capture of K$^-$.
The energy range was extended to $E_{\gamma}^{\rm L} <$ 2.1 GeV.
The C4, however, violates the SU(3) symmetry constraint for both
the main coupling constants: $g_{\rm K\Lambda N}/\sqrt{4\pi} =
-2.38$ and $g_{\rm K\Sigma N}/\sqrt{4\pi} = 0.27$. The model SLA
focused to description of the data in the energy range up to 2.5
GeV. The SU(3) limits for the coupling constants are fulfilled
since they were included in the fitting procedure. More details
and comparison of the models can be found in Ref.~\cite{prep03}.

\begin{figure}[htb]
\begin{center}
  \includegraphics[width=8.6cm]{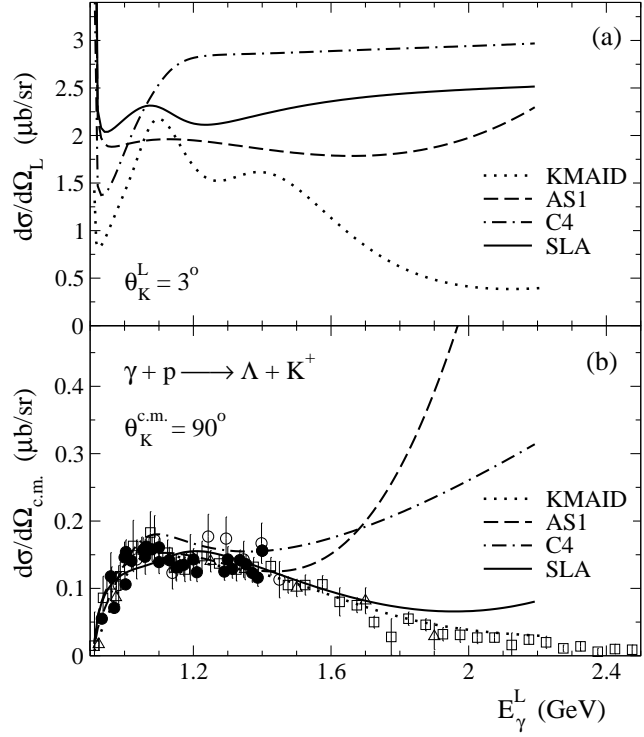}
\end{center}
\vspace*{-5mm} \caption{ Laboratory (a) and center of mass (b)
cross sections are plotted as a function of the photon laboratory
energy at fixed kaon angles. Predictions of the models KMAID, AS1,
C4, and SLA (see the text) are compared with experimental data. In
the laboratory frame (a) only the theoretical curves are shown.
The data are from Refs.~\cite{exp} (solid circles),~\cite{SPH94}
(circles),~\cite{SPH98} (triangles), and~\cite{SPH03} (squares).}
\label{figure3}
\end{figure}

%
% Figure 4
%
In Figure~\ref{figure3} we compare differential cross sections as
they are predicted by the four models with experimental data at
fixed kaon angle as a function of the photon laboratory energy. In
the bottom part (b), the c.m. cross section is shown at kaon c.m.
angle of 90$^\circ$. Predictions of all models are in a very good
agreement with the data for energies up to 1.4 GeV but for the
higher energies the models AS1, and C4 overestimate significantly
the SAPHIR data~\cite{SPH98} which were not included in the
fitting procedure of neither of them. Only the KMAID and SLA
models are consistent with the data up to 2 GeV. All the four
models provide acceptable description of the process at
$E_{\gamma}^{\rm L}\approx 1.3$ GeV concerned in the present
hypernuclear calculations. The amount of the cross section
difference coming from these different models constitutes a part
of theoretical uncertainty in predicting the hypernuclear
production rate.

However, in the hypernuclear calculations only the forward angle
laboratory amplitude for the elementary process enters effectively
into the calculations. In Figure~\ref{figure3}(a) we show behavior
of theoretical laboratory cross sections at $\theta_{\rm K}^{\rm
L}=3^\circ$. Predictions of the models AS1, C4 and SLA reveal a
constant discrepancy in most of the energy range whereas the KMAID
model predicts considerably different values with respect to the
SLA near the threshold and especially at energies lager than 1.6
GeV. At 1.3 GeV and $\theta_{\rm K}^{\rm L}=3^\circ$, SLA gives by
40\% larger cross section than KMAID.

\begin{figure}[htb]
\begin{center}
  \includegraphics[width=8.6cm]{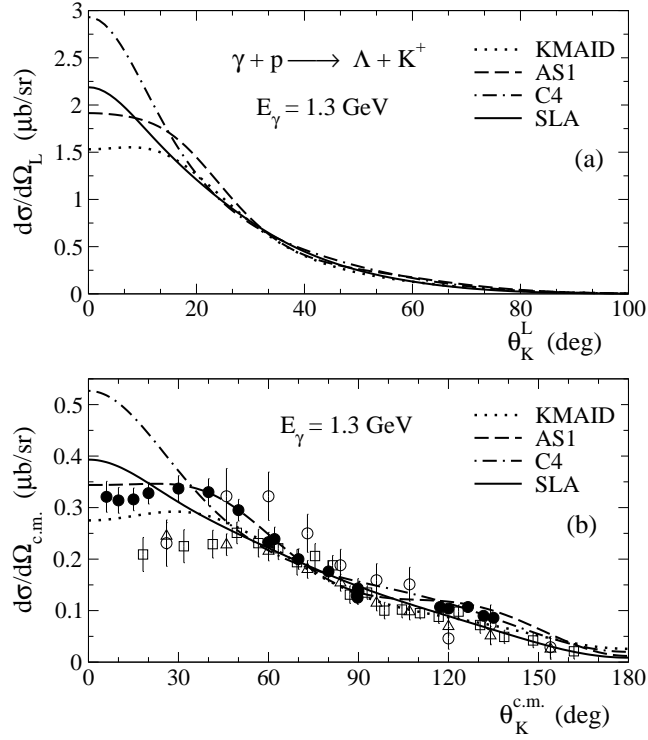}
\end{center}
\caption{The same as in Fig.~\ref{figure3} but for the angular
dependence at $E^{\rm L}_{\gamma}=1.3$ GeV.}
\label{figure4}
\end{figure}

%
% Figure 5
%
In Figure~\ref{figure4}  we plot angular distribution of the cross
sections at  $E^{\rm L}_{\gamma}=1.3$ GeV. Predictions of the
models are in a good agreement with the data (the bottom part (a))
for all angles except for $\theta_{\rm K}^{\rm c.m.} <$ 30$^\circ$
where C4 over-predicts the data. Note here, probably a systematic,
discrepancy of SAPHIR data~\cite{SPH98,SPH03} and those by
Bleckmann {\it et al.}~\cite{exp} for $\theta^{\rm c.m.}_{\rm K}
<$ 45$^\circ$. The latest CLAS measurements~\cite{CLAS} confirm
the discrepancy between the data sets which makes fixing the
theoretical models at small angles problematic. In
Figure~\ref{figure4}(a) large differences between results of the
models at forward angles are demonstrated showing that an
uncertainty can amount up to 80\% (KMAID and C4) at very forward
angles ($\theta_{\rm K}^{\rm L}$ = 3$^\circ$).
Figure~\ref{figure3}(a) shows that the C4 predicts too large cross
sections at small angles in the whole energy range whereas KMAID
drops down at $E^{\rm L}_{\gamma}> 1.6$ GeV making the difference
still larger.

\begin{figure}[htb]
\begin{center}
  \includegraphics[width=8.6cm]{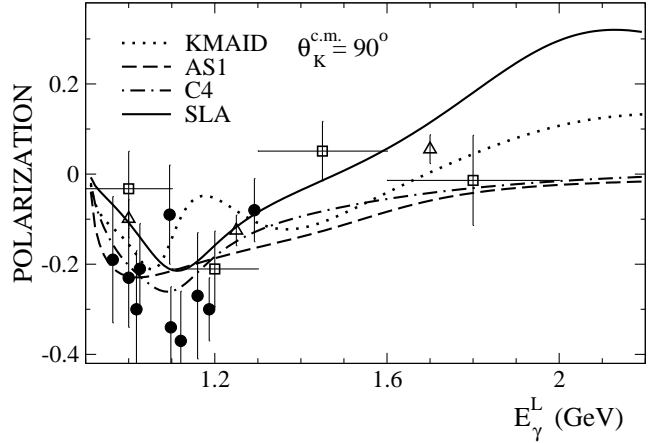}
\end{center}
\caption{ Polarization of $\Lambda$ is plotted as a function of
the photon laboratory energy at 90$^\circ$ in the $\gamma {\rm
p}\rightarrow \Lambda {\rm K}^+$ reaction. Results of the four
theoretical models (see the text) are compared with experimental
data. The data are from Refs.~\cite{exp} and~\cite{expP} (solid
circles),~\cite{SPH98} (triangles), and~\cite{SPH03} (squares).}
\label{figure5}
\end{figure}

\begin{figure}[htb]
\begin{center}
  \includegraphics[width=8.6cm]{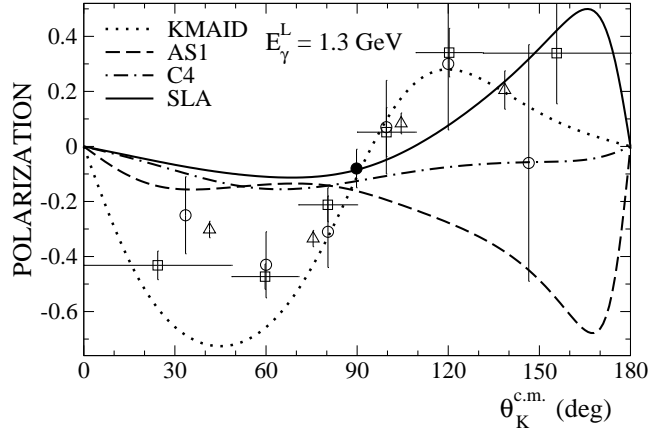}
\end{center}
\caption{The same as in Fig.~\ref{figure5} but for the angular
distribution at $E^{\rm L}_{\gamma}$ = 1.3 GeV. The data are from
Refs.~\cite{exp} (solid circles),~\cite{SPH94} (empty circles),~
\cite{SPH98} (triangles), and~\cite{SPH03} (squares).}
\label{figure6}
\end{figure}

%
% Figure 6
%
The most remarkable difference between predictions of the models
is revealed when the hyperon polarization is plotted. In
Figure~\ref{figure5} we compare the $\Lambda$ polarization at
$\theta_{\rm K}^{\rm c.m.}$ = 90$^\circ$ as a function of photon
laboratory energy. The quality of the data, however, does not
allow to prefer some of the models. This is not the case of the
angular dependence of the polarization shown in Fig.~\ref{figure6}
where the results of the models exhibit significant deviations at
the backward angles. Prediction of the AS1 model is out of the
data at large angles having opposite sign to that of the data
points. However, all the four models seem to have a problem with
description of the data points around 60$^\circ$. Of the four
models the KMAID and SLA provide large values of the polarizations
at backward angles with the proper sign. These models deviate also
from the others for $E_{\gamma}^{\rm L}\approx 2$ GeV at
90$^\circ$ (Fig.~\ref{figure5}).
%
% Figure 7
%

In order to give an idea of relative importance of the
spin-non-flip ($f_0$) and spin-flip ($g_0$, $g_{1}$, $g_{-1}$)
amplitudes, in Table~\ref{tab31gk} we list values of the
amplitudes in the laboratory frame calculated for $\theta_{\rm
K}^{\rm L} = $3$^\circ$ and 10$^\circ$ at $E_{\gamma}^{\rm L}=
1.3$ GeV. The amplitudes are normalized as follows
\begin{equation}
\left(\frac{d\sigma}{d\Omega}\right)_{\rm L}= \frac{s\: p_{\rm
K}^2\, E_{\rm K}\, E_{\Lambda}} {p_{\rm K}(E_{\Lambda}+E_{\rm
K})-E_{\gamma}E_{\rm K}\cos\theta_{\rm K}^{\rm L}}\:
\frac{1}{2}\,(\,|f_0|^2+|g_0|^2+|g_1|^2+|g_{-1}|^2\,)\,,
\end{equation}
where the energies, momenta, and angle are in the laboratory
frame. It is evident that at small scattering angles which are
relevant in hypernuclear calculations the three spin-flip
amplitudes are much larger in magnitude than the spin-non-flip
one. This observation together with the high momentum transfer to
$\Lambda$ (Fig.~\ref{figure2}) suggest that a large number of
possible hypernucleus final states with high spins and unique
features can be selectively populated.
%
% Table 1
%

An extension of the isobaric models to the electro-production
process is carried out by introduction of a $Q^2$-dependence into
the electro-magnetic vertices by means of form factors~
\cite{WC92,SL96}. However, in the calculations of the
electro-production of hypernuclei, performed here (small kaon and
electron angles), very small values of $Q^2$ [$Q^2 < 0.1$
(GeV/$c$)$^2$] are reached which, due to a smooth $Q^2$-dependence
of the form factors, results in a very small effect in the
amplitudes. It means that the elementary photo-production
amplitude AS1 can be also utilized for the calculations of the
electro-production of hypernuclei at small angles. This is also
why we did not discuss results of the models for the
electro-production here but it can be found in Ref.~\cite{prep03}.

%
%  Section 4
%
\section{Excitation spectra predicted for
$^{28}{\rm Si}(\gamma,{\rm K}^+)^{28}_{\Lambda}{\rm Al}$ }

In order to demonstrate characteristics of photo-production of
hypernuclei, here we choose $^{28}$Si as a typical nuclear target.
The excitation spectra have been evaluated at $E^{\rm
L}_{\gamma}=1.3$ GeV and $\theta^{\rm L}_{\rm K}=3^\circ$ which
are close to the kinematical condition in the experimental
proposal. In the first subsection, for demonstration the $^{28}$Si
target ground state is assumed to have the lowest proton-closed
shells $[s^4p^{12}(0d_{5/2})_{\rm pn}^{12}]$. In the second
subsection, we extend the calculation to employ realistic wave
functions solved in the full $[s^4p^{12}(sd)^{12}_{\rm pn}]$
shell-model space.

%
%  Subsection 4.1
%
\subsection{A demonstrative model with $(0d_{5/2})^6_{\rm p}$ }

As the proton shells are closed up to $(0d_{5/2})_{\rm p}^6$, the
final hypernuclear states are described, respectively, with 1p-1h
configurations $[(nlj)_{\rm p}^{-1}(nlj)^{\Lambda}]_J$. For the
single-particle wave functions, we employ the DDHF solutions so as
to be as realistic as possible. The calculated cross sections are
summarized in Table~\ref{Tab:Si28d5}.
%
% Table 2
%

The characteristic result to be emphasized first is the selective
excitation of the highest-spin state within each 1h-1p multiplet.
In fact one sees in Table~\ref{Tab:Si28d5} that the
$[d_{5/2}^{-1}\,s_{1/2}^{\Lambda}]_{3^+}$,
$[d_{5/2}^{-1}\,p_{3/2}^{\Lambda}]_{4^-}$,
$[d_{5/2}^{-1}\,p_{1/2}^{\Lambda}]_{3^-}$,
$[d_{5/2}^{-1}\,d_{5/2}^{\Lambda}]_{5^+}$, and
$[d_{5/2}^{-1}\,d_{3/2}^{\Lambda}]_{4^+}$ states are very strongly
excited and that the cross sections to the lower-spin states are
much smaller. The preferential excitation of the high-spin states
are attributed to the large momentum transfer (about 350 MeV/$c$)
as similarly as in the case of the $(\pi^+,{\rm K}^+)$ reaction.

Secondly, such a novel fact is revealed that the selectively
excited state in each combination $[j_>^{-1}j_>^{\Lambda}]_J$ has
an unnatural parity with the maximum spin value of $J=J_{\rm
max}=j_>+j_>^{\Lambda}=l_{\rm p}+l_{\Lambda}+1=L_{\rm max}+1$. In
Table~\ref{Tab:Si28d5}, one may refer to the
$[d_{5/2}^{-1}\,s_{1/2}^{\Lambda}]_{3^+}$,
$[d_{5/2}^{-1}\,p_{3/2}^{\Lambda}]_{4^-}$, and
$[d_{5/2}^{-1}\,d_{5/2}^{\Lambda}]_{5^+}$  cases for
reconfirmation. The
$[p_{3/2}^{-1}\,j_>^{\Lambda}]_{J=2^-,3^+,4^-}$ states in the
lower block of Table~\ref{Tab:Si28d5}  have the similar nature.
This kind of selectivity is not seen in the other hypernuclear
production processes  such as   $(\pi^+,{\rm K}^+)$ and $({\rm
K}^-,\pi^-)$ reactions. This is attributed to the spin-flip
transition dominance in the elementary hyperon photo-production
reaction (see Tab.~\ref{tab31gk}). It is also noted that, in the
other combinations such as $[j_>^{-1}j_<^{\Lambda}]_J$ or
$[j_<^{-1}j_>^{\Lambda}]_J$, the highest spin is limited to
$J'_{\rm max}=l_{\rm p}+l_{\Lambda}$ and accordingly the natural
parity to $(-1)^{l_{\rm p}+l_{\Lambda}}$.

\begin{figure}[htb]
  \begin{center}
   \includegraphics[height=8.6cm,angle=270]{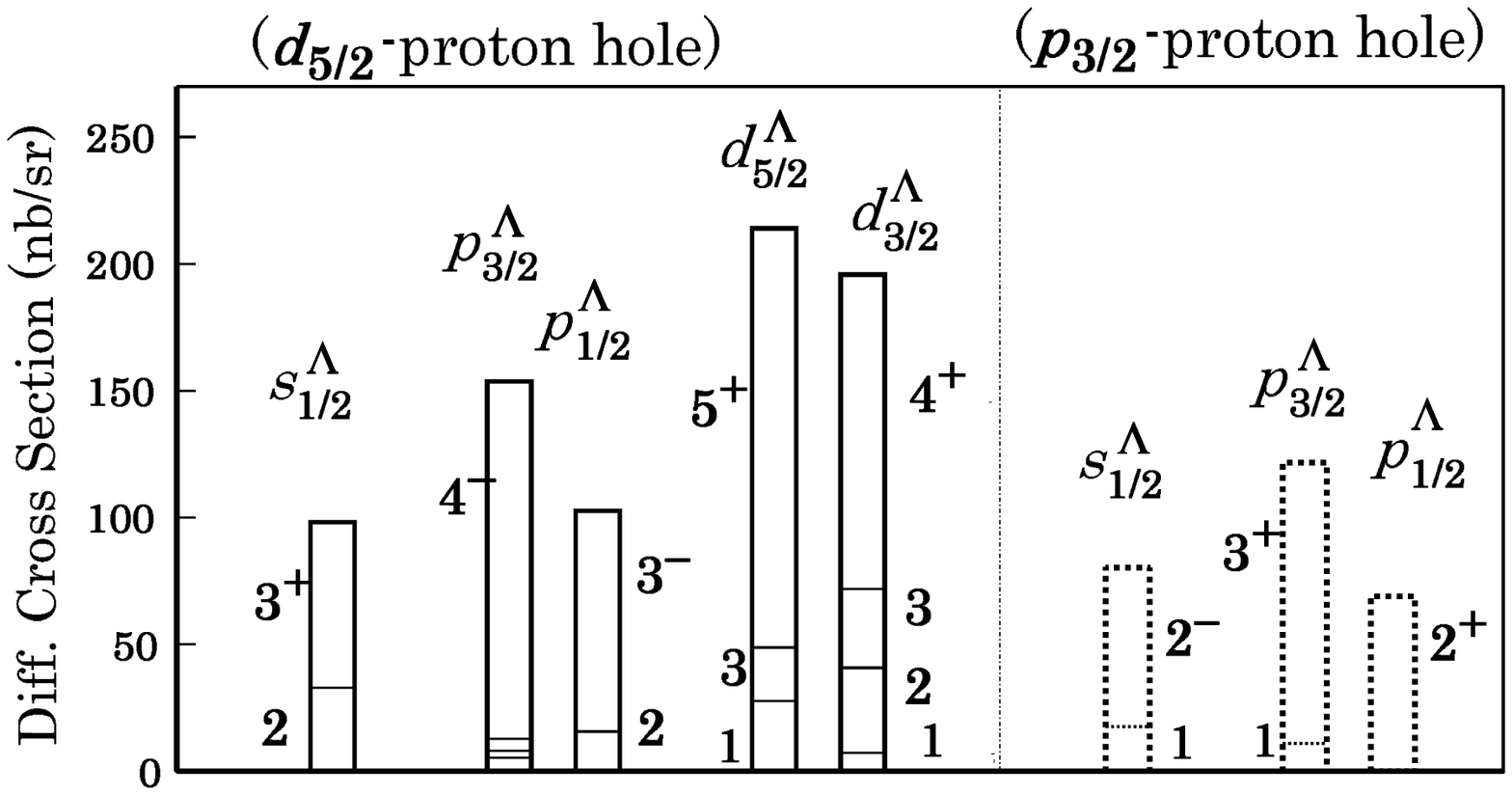}
  \end{center}
\caption{Divided contributions to the particle-hole $J$-multiplet
state $[j_{\rm p}^{-1}j^{\Lambda}]_J$ as calculated for the
$^{28}$Si($\gamma,{\rm K}^+)^{28}_{\,\Lambda}$Al reaction at
$E_{\gamma}=1.3$ GeV and $\theta_{\rm K}^{\rm L}=3^\circ$. The
DDHF wave functions are used with the Saclay-Lyon A
amplitude~\cite{SLA98}. The height of each pillar corresponds to
the differential cross section, while the width has no special
meaning.} \label{fig:Si28gKJmax}
\end{figure}

The numerical results of Table~\ref{Tab:Si28d5} are schematically
shown in Fig.~\ref{fig:Si28gKJmax} where relative strengths
in each J-multiplet are easily understood.
%
% Figure 8
%

\begin{figure}[htb]
  \begin{center}
  \includegraphics[height=7.6cm]{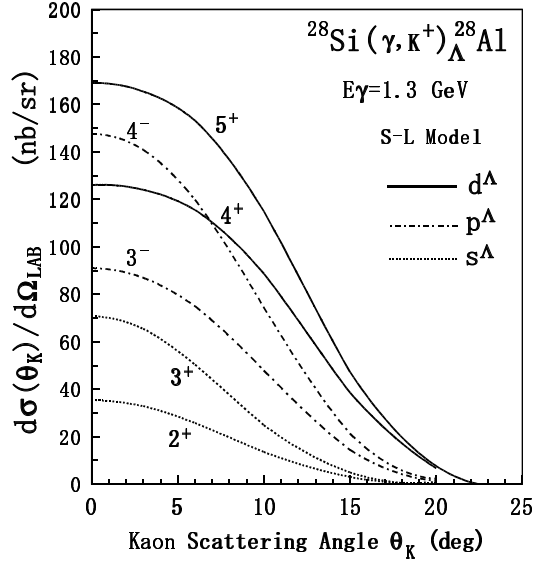}
  \end{center}
\caption{Calculated angular distributions of the states
excited strongly in $^{28}$Si($\gamma,{\rm K}^+)^{28}_{\,\Lambda}$Al
reaction at  $E_{\gamma}=1.3$ GeV. The excited states
denoted with $J$ correspond to the $[j_p^{-1}j^{\Lambda}]_J$
multiplets shown in Fig.~\ref{fig:Si28gKJmax}. }
\label{fig:AngDist28}
\end{figure}

Figure~\ref{fig:AngDist28} shows the calculated angular distributions
for the pronounced peaks. All these differential cross sections
decrease quickly as the kaon lab scattering angle increases. It is
interesting to note that the relative strength for the
$[d_{5/2}^{-1}\,d_{3/2}^{\Lambda}]_{J=4^+}$ and
$[d_{5/2}^{-1}\,p_{3/2}^{\Lambda}]_{J=4^-}$ states changes
 at $\theta_{\rm K}^{\rm L}\simeq 7^\circ$.
%
% Figure 9
%

%
%  Subsection 4.2
%
\subsection{Use of the $(sd)^n$ full space wave function}

Here we use sophisticated wave functions solved in the full
$(0d_{5/2}\,0d_{3/2}\,1s_{1/2})_{\rm pn}^{11,12}$ space for
$^{27,28}$Si. It is remarked that the use of such detailed wave
functions should predict several new but minor states in addition
to the pronounced peaks which are expected in the simplified
configuration adopted in the preceding subsection. The present
estimates are based on the spectroscopic amplitudes for proton
pick-up from $^{28}$Si leading to the excited states of $^{27}$Al
which are calculated with the $(sd)^n$ model space.

\begin{figure}[htb]
  \begin{center}
  \includegraphics[height=8.6cm,angle=270]{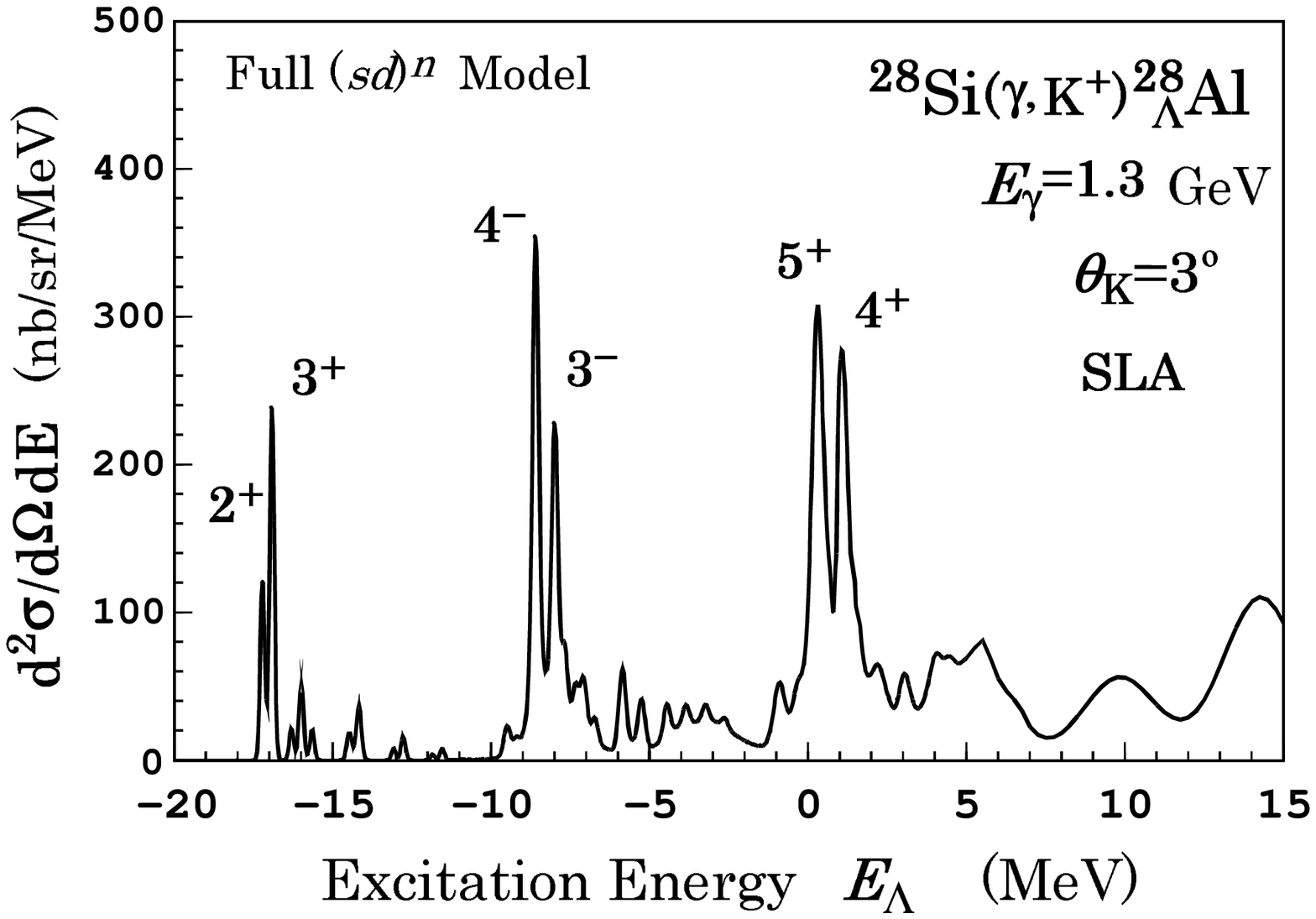}
  \end{center}
\caption{Theoretical excitation function calculated with
the full $(sd)^n$ wave functions for the
$^{28}$Si($\gamma,{\rm K}^+)^{28}_{\,\Lambda}$Al reaction at
 $E_{\gamma}=1.3$ GeV and $\theta_{\rm K}^{\rm L}=3^\circ$
using the Saclay-Lyon A model. For simplicity to draw pronounced
doublet peaks, the artificial $ls$-splitting is introduced as
0.17(2$l$+1) in MeV, so that in actual case such multiplet may be
seen as a degenerate one. The hypernuclear energy is measured from
the $^{27}$Al(g.s.)+$\Lambda$ threshold, so it is expressed in
terms of the hyperon energy $E_{\Lambda}$.} \label{fig:A28fullsf}
\end{figure}

In order to predict a realistic excitation spectrum for the
$^{28}$Si($\gamma,{\rm K}^+)^{28}_{\,\Lambda}$Al reaction, one has
to take the empirical proton-hole widths into account, although
they are not always available.  Figure~\ref{fig:A28fullsf} shows
the result where the following proton widths are employed
tentatively: $\Gamma_{\rm p}(0s_{1/2}^{-1})=10$ MeV, $\Gamma_{\rm
p}(0p_{3/2}^{-1})=6$ MeV, $\Gamma_{\rm p}(0p_{1/2}^{-1})=3$ MeV,
and $\Gamma_{\rm p}(0d_{5/2}^{-1})=0$ MeV. At the same time, for
the $\Lambda$ bound states the width $\Gamma_{\Lambda}(j)=0.3$ MeV
is used which is about half of the energy resolution expected at
the Jefferson Lab, while  $\Gamma_{\Lambda}(j)=1.0$ MeV for
$0<E_{\Lambda}<2$ MeV and $\Gamma_{\Lambda}(j)=3$ MeV for
$E_{\Lambda}>2$ MeV are assumed rather arbitrarily. Furthermore
the energy splittings between members of the  $[j_{\rm
p}^{-1}j_<^{\Lambda}]_J$ multiplet are taken from the YNG($\Lambda
{\rm N}$) h-p interaction~\cite{Yamamoto117} derived from the
Nijmegen model-D, and it is notable that the splittings are mostly
of the order of 0.1 MeV.
%
% Figure 10
%

Major 3 doublets (6 peaks) structure obtained  with the simplified
wave functions $[d_{5/2}^{-1}\,j^{\Lambda}]$ (see Table 2) well
persist also in the new estimates. It is quite interesting to note
that, for the major peaks, the use of the full space wave
functions results in the reduction of the cross sections by a
factor of about 0.65 in comparison with the single-$j$ estimate
with $(0d_{5/2})_{\rm p}^6$. It should be also remarked that two
pronounced peaks obtained at  $E_{\Lambda}\approx -8.5$ MeV
correspond to the $[d_{5/2}^{-1}\,p_{3/2}^{\Lambda}]_{4^-}$ and
$[d_{5/2}^{-1}\,p_{1/2}^{\Lambda}]_{3^-}$ structure, respectively.
As the hole-particle interactions for high-spin states are
generally very small, the energy difference between these two
peaks, if separated experimentally, provides us the spin-orbit
splitting of the $\Lambda$ $p$-state.

The third major doublet obtained at  $E_{\Lambda}\approx 0$~MeV
includes the unnatural parity highest-spin state
 $[d_{5/2}^{-1}\,d_{5/2}^{\Lambda}]_{5^+}$ which should
get the biggest cross section. If the
$d_{5/2}^{\Lambda}$ state is bound or it is not bound but
the energy position is not so high above the threshold,
this peak width might be sharp enough to be identified  in
the experiment with the good energy resolution expected at
JLab. As the partner has the dominant structure of
 $[d_{5/2}^{-1}\,d_{3/2}^{\Lambda}]_{4^+}$ and the
 $\Lambda$N hole-particle interactions in high-spin
states are very small, the energy difference between
these two peaks is almost equal to the spin-orbit splitting
of the $d$-state $\Lambda$. Thus the photo- and electro-production
reactions will provide a nice opportunity of looking at such
splittings in heavy systems if the energy resolution is good
enough.

It is interesting to see how different the theoretical
hypernuclear production cross sections are when one
uses different models of hyperon photo-production
amplitudes. In section 3 we have already discussed
remarkable difference among the AS1, KMAID, C4, and
SLA models. In Table~\ref{Tab:Si28Al} the cross sections
of strongly excited states in
$^{28}$Si($\gamma,{\rm K}^+)^{28}_{\,\Lambda}$Al are compared
as calculated with the SLA and KMAID amplitudes. They are
the cross sections expected within the $(sd)^n$ shell
model framework, noting that use of the $(sd)^n$ wave
functions give rise to a reduction factor of 0.61
in comparison with the single-j estimate of
$(d_{5/2})^n$. One notes from Table~\ref{Tab:Si28Al}
that the KMAID amplitude gives approximately 30\%
smaller cross sections of the SLA values when compared
at $\theta^{\rm L}_{\rm K}=3^\circ$. Such difference has
been suggested already in section 3.
%
% Table 3
%

The present treatment is a direct extension of that for
the $p$-shell hypernuclear photo-production
calculation~\cite{SF,MotobaPTP117} where always the full
$p$-shell wave functions have been easily employed.
 In the case of $p$-shell hypernuclear
production, the particles involved in the low-lying state are in
the $s$- and $p$-orbits, so that the ``high-spin'' selectivity
mentioned above is realized as the transitions with $\Delta
J=\Delta L+\Delta S=2^-$, $2^+$, and $3^+$. It is worthwhile to
remark that in Fig.~\ref{fig:A28fullsf} there appear side peaks at
$E_{\Lambda}\simeq -16$ and $-$14 MeV in $^{28}_{\,\Lambda}$Al as
similarly as confirmed in $^{12}_{\,\Lambda}$C ($1_2^-$ and
$1_3^-$ ). They are based on the $s$-state $\Lambda$ particle
coupled to the excited states ($3/2^+$ and $7/2^+$) in $^{27}$Al.

%
%  Section 5
%
\section{Photo-production with the $^{40}{\rm Ca}$ and
$^{52}{\rm Cr}$ targets}

\begin{figure}[htb]
  \begin{center}
  \includegraphics[height=8.6cm,angle=270]{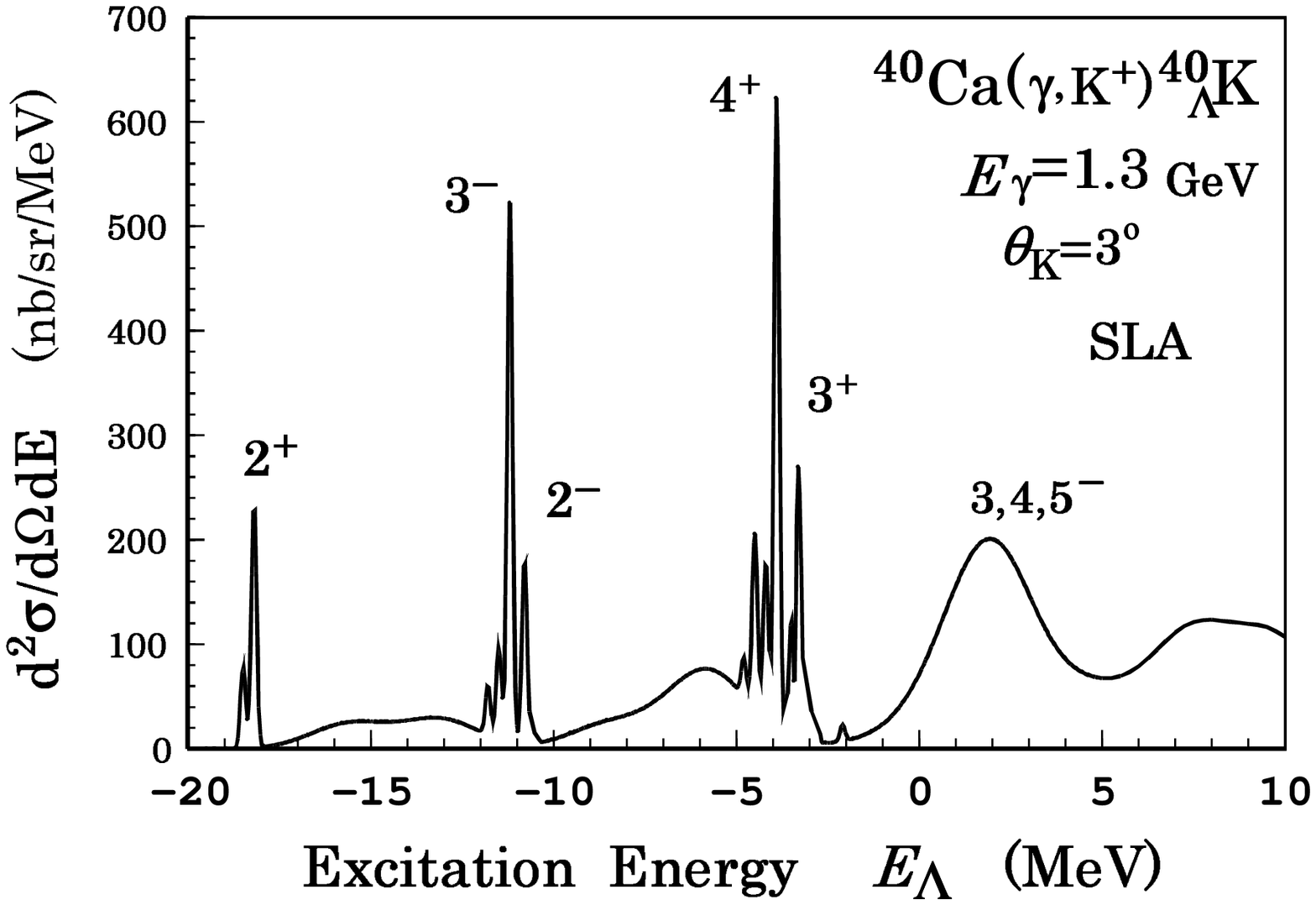}
  \end{center}
\caption{ Excitation function for the $^{40}$Ca($\gamma,{\rm
K}^+)^{40}_{\,\Lambda}$K reaction calculated in DWIA at
$E_{\gamma}=1.3$ GeV and $\theta^{\rm L}_{\rm K}=3^\circ$ using
the Saclay-Lyon A model. For simplicity to draw pronounced doublet
peaks, the artificial $ls$-splitting is introduced as 0.17(2$l$+1)
in MeV. In actual case such multiplet may be seen as a degenerate
one. The hypernuclear energy is measured from the
$^{39}$K(g.s.)+$\Lambda$ threshold, so it is expressed in terms of
the hyperon energy $E_{\Lambda}$.} \label{fig43gk}
\end{figure}

The next sample target with heavier mass is $^{40}$Ca and the
calculated excitation function for the $^{40}$Ca($\gamma,{\rm
K}^+)^{40}_{\,\Lambda}$K is shown in Fig.~\ref{fig43gk}. This
target is doubly $LS$-closed up to the $0d_{3/2}$ shell, so that
the situation is different from the $^{28}$Si case because
$^{40}$Ca has the uppermost proton orbit of the $j_<\,$-type.
Therefore the highest spin in a $[d_{3/2}^{-1}\,j_>^{\Lambda}]_J$
multiplet is $J'_{\rm max}=j_<+j_>^{\Lambda}=l_{\rm
p}+l_{\Lambda}=L_{\rm max}$ with a natural parity. On the other
hand, in a $[d_{3/2}^{-1}\,j_<^{\Lambda}]_J$ multiplet the highest
spin is $J''_{\rm max}=j_<+j_<^{\Lambda}=l_{\rm
p}+l_{\Lambda}-1=L_{\rm max}-1$, so that this restriction on the
smaller angular momentum transfer makes the latter cross sections
much smaller than the former case with $J'_{\rm max}=L_{\rm max}$.
This situation of the $j_<\,$-closed shell explains why the
dominant peak series shown in Fig.~\ref{fig43gk} consists of the
natural parity $2^+$, $3^-$, and $4^+$ states accompanied with
only minor side peaks of the $1^+$, $2^-$, and $3^+$ states,
respectively. In other word, if these pronounced peak positions
are measured with good energy resolution in a future experiment,
they tells us the exact energies of the $\Lambda$ particle in the
$s_{1/2}^{\Lambda}$, $p_{3/2}^{\Lambda}$, and $d_{5/2}^{\Lambda}$,
respectively. The broad background shown in Fig.~\ref{fig43gk}
reflects the strengths originating from the deeper proton shell,
$0d_{5/2}$, which has the spreading width of several MeV. In
reality we also expect extra small peaks between the pronounced
peaks, as they might be based on the fragmentation of the
$0d_{3/2}$ proton hole strength. In spite of such factors, the
photo-production reaction with this kind of doubly $LS$-closed
target might give a nice example of showing up the $\Lambda$
single-particle energies in medium-heavy mass region.
%
% Figure 11
%

\begin{figure}[htb]
  \begin{center}
  \includegraphics[height=8.6cm,angle=270]{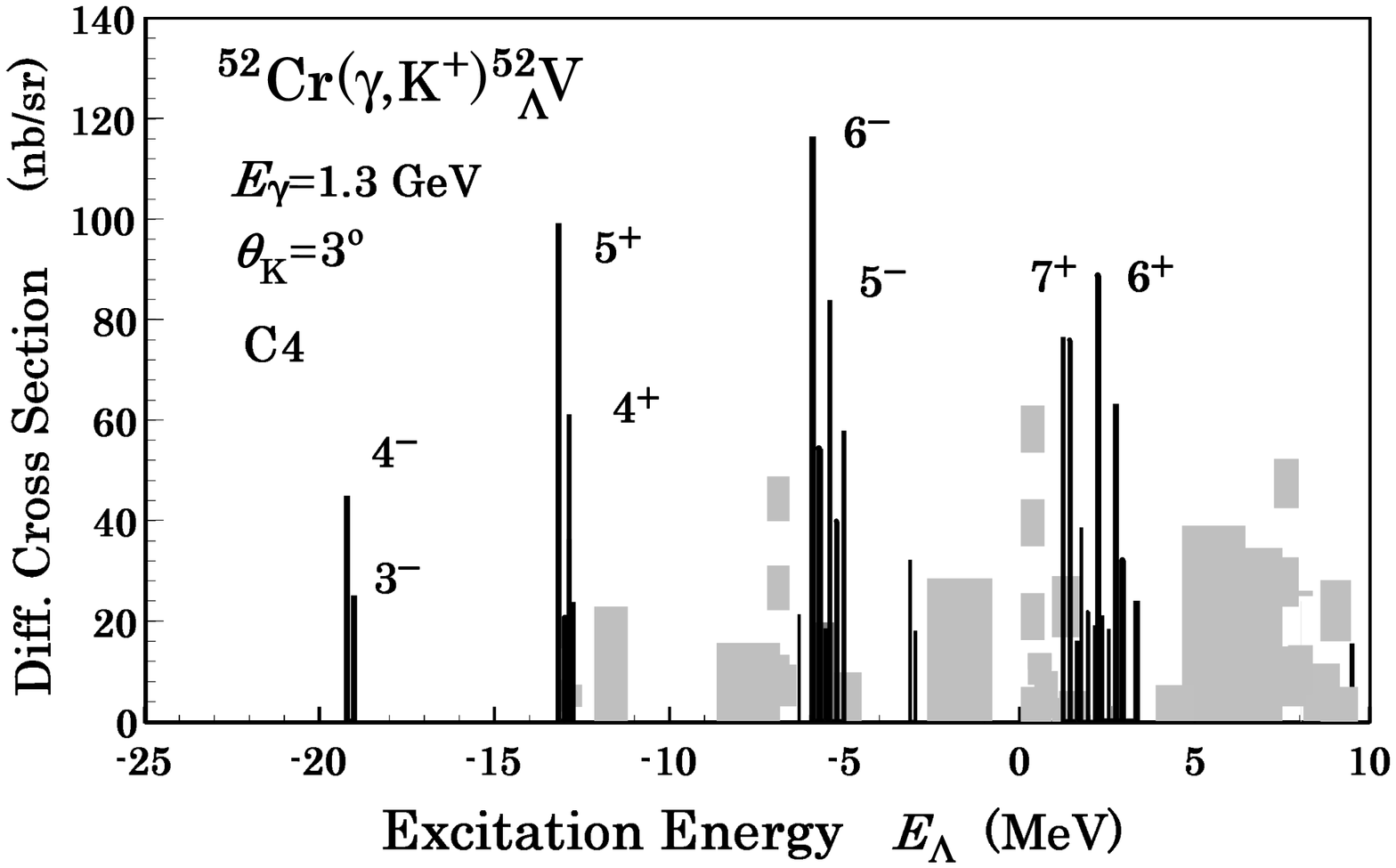}
  \end{center}
\caption{ Excitation function for the $^{52}$Cr($\gamma,{\rm
K}^+)^{52}_{\,\Lambda}$V reaction calculated in DWIA at
$E_{\gamma}=1.3$ GeV and $\theta^{\rm L}_{\rm K}=3$ deg using the
C4 model. For simplicity to draw pronounced doublet peaks, the
artificial $ls$-splitting is introduced as 0.17(2$l$+1) in MeV. In
actual case such multiplet may be seen as a degenerate one. See
text for gray blocks. The hypernuclear energy is measured from the
$^{51}$V(g.s.)+$\Lambda$ threshold, so it is expressed in terms of
the hyperon energy $E_{\Lambda}$.} \label{fig52gk}
\end{figure}

Finally we add the case of the $^{52}$Cr target as a typical
candidate from the $fp$-shell region of nuclei. The calculated
results for the $^{52}$Cr($\gamma,{\rm K}^+)^{52}_{\,\Lambda}$V
reaction spectrum is displayed in Fig.~\ref{fig52gk}, where the
vertical axis shows simply the cross section value in nb/sr.
Therefore, differently from the former figures, here we do not
take the smearing width into account when we show the major series
of the pronounced peaks by solid lines. The C4 elementary
amplitude is employed here, although it leads to overestimates of
the cross section by about 25\%. It is also noted that in
Fig.~\ref{fig52gk} we show positions of broad peaks by gray blocks
originating from deeper proton hole orbits such as
$(0d_{3/2}^{-1})_{\rm p}$ and $(0d_{5/2}^{-1})_{\rm p}$. The
height$\times$width of each gray block correspond to the
calculated cross section estimates
%
% Figure 12
%

The $^{52}$Cr target has four protons in the uppermost
$j_>=0f_{7/2}$ shell and the neutron is $jj$-closed. Therefore it
is easy to understand that the major peak series are based on the
conversion of the $0f_{7/2}$ protons into the $\Lambda$-particle
sitting in the $s$, $p$, $d$, and $f$ orbits. In fact we confirmed
by the calculation that the dominant peak series consists of the
unnatural parity $[f_{7/2}^{-1}\,j_>^{\Lambda}]_{J=J_{\rm max}}$
states with $J_{\rm max}=j_>+j_>^{\Lambda}=l_{\rm
p}+l_{\Lambda}+1=L_{\rm max}+1$ where
$j_>^{\Lambda}=s_{1/2}^{\Lambda}(J=4^-)$,
$p_{3/2}^{\Lambda}(J=5^+)$, $d_{5/2}^{\Lambda}(J=6^-)$, and
$f_{7/2}^{\Lambda}(J=7^+)$, respectively. On the other hand, the
$\Lambda$ spin-orbit partner states
$[f_{7/2}^{-1}\,j_<^{\Lambda}]_{J=L_{\rm max}}$ ($J=3^-, 4^+, 5^-,
6^+$) gain about 60\% production rate of the corresponding biggest
peak within each multiplet. Therefore, if the energy resolution is
good enough, there will be a chance to get information on the
$ls$-splitting in this medium-mass region. In the present case,
the other angular momentum states also have certain (non
negligible) contributions to each peak. It should be mentioned
that, if we improve the description of the $^{51}$V nuclear
excited states, we will also get very weak excited side peaks
among the strong peaks shown here. See also the
report~\cite{Sotona03}.

\section{Conclusion}

The first (e,e'K$^+$) experiment on the nuclear target ($^{12}$C)
was reported recently from the Jefferson Lab, proving it to be a
nice tool for spectroscopic study of $\Lambda$ hypernuclei. As our
theoretical prediction has been confirmed remarkably by this
experiment, we applied its framework to typical heavier targets
with careful consideration of modern elementary amplitudes for
hyperon photo-production. In this paper we presented the
($\gamma$,${\rm K}^+$) excitation spectra for producing
medium-mass $\Lambda$-hypernuclei, where we focused our attention
to the novel and characteristic features of the reaction process.

We started with discussions on the basic properties of modern
elementary amplitudes of kaon photo-production process. Among many
theoretical attempts, the isobaric models for the $ \gamma {\rm p}
\to \Lambda {\rm K}^+$ process based on the Feynman diagram
technique are of interest and successful in the kinematical region
assumed here, $E_\gamma^{\rm L} = 0.91 - 2.0$ GeV. Four elementary
amplitudes of the models are adopted in this paper, i.e., those
denoted as Kaon-MAID (KMAID), Adelseck-Saghai (AS1),
Williams-Ji-Cotanch (C4) and Saclay-Lyon A (SLA) models. The
models differ in their own choices of nucleon and hyperon
resonances. Detailed comparison of the models has been done for
the differential cross sections and polarizations for the
elementary $\gamma {\rm p} \to \Lambda {\rm K}^+$ and their
results are compared with the available experimental data. Some
variances of agreement between the models exist at higher energies
$E_\gamma^{\rm L} \ge 1.4$ GeV. However, all the four models are
acceptable at $E_\gamma^{\rm L} \approx 1.3$ GeV which is
concerned in the present hypernuclear calculations. The unique and
interesting feature of the $ \gamma {\rm p} \to \Lambda {\rm K}^+$
amplitudes lies in that the spin-flip amplitudes are much larger
in magnitude than the spin-non-flip one even at the forward angles
at $\theta^{\rm L}_{\rm K} = 3^\circ$ and $\theta^{\rm L}_{\rm K}
= 10^\circ$ and at energy $E_\gamma^{\rm L} = 1.3$ GeV. This
feature of the amplitude together with the large momentum transfer
($q_\Lambda \approx 350-400$ MeV/$c$) yields the novel and
selective features of the excitation spectra in the medium-heavy
$\Lambda$-hypernuclei.

Here we choose $^{28}{\rm Si}$, $^{40}{\rm Ca}$ and $^{52}{\rm
Cr}$ as the typical medium-mass target nuclei for the
spectroscopic study of the $\Lambda$-hypernuclei in the $(\gamma,
{\rm K}^+)$ reactions. First the excitation spectra of $^{28}{\rm
Si}(\gamma,{\rm K}^+)^{28}_{\,\Lambda}{\rm Al}$ are discussed for
two model calculations of $^{28}{\rm Si}$. When the simple
$(0d_{5/2})_{\rm p}^6\,$-closed configuration is assumed for
$^{28}{\rm Si}$ as a demonstrative model, the characteristic and
unique excitation function is clearly seen. The result emphasized
first is the selective excitation of the highest-spin state within
each 1h-1p multiplet such as
$[d_{5/2}^{-1}\,p_{3/2}^\Lambda]_{4^-}$,
$[d_{5/2}^{-1}\,p_{1/2}^\Lambda]_{3^-}$,
$[d_{5/2}^{-1}\,d_{5/2}^\Lambda]_{5^+}$ and
$[d_{5/2}^{-1}\,d_{3/2}^\Lambda]_{4^+}$.  This is due to the large
momentum transfer for the reaction process. Second, such a novel
fact is revealed that the selectively excited states with
$[j_{>}^{-1} j_{>}^\Lambda]_J$ are strongly populated and have
unnatural parities with maximum spins of $J = J_{\rm max} = j_{>}
+ j_{>}^\Lambda = \ell_{\rm p} + \ell_\Lambda +1 = L_{\rm max} +
1$.  This is attributed to the spin-flip transition dominance in
the elementary hyperon photo-production reaction. This kind of
selectivity is not seen in other hypernuclear production processes
such as $({\rm K}^-,\pi^-)$ and $(\pi^+,{\rm K}^+)$ reactions.

The calculation has been extended to the $(sd)^n$ full space model
to describe $^{28}{\rm Si}$. Excitation function for the
$^{28}{\rm Si}(\gamma,{\rm K}^+)^{28}_{\,\Lambda}{\rm Al}$ is
presented, taking the empirical particle-hole width into account
tentatively so as to make the comparison with the future
experiment possible. The major doublets and peaks obtained with
the simplified model well persist also in the extended estimates.
The energy splitting of the two peaks with structures, such as
$[d_{5/2}^{-1}\,p_{3/2}^\Lambda]_{4^-}$ and
$[d_{5/2}^{-1}\,p_{1/2}^\Lambda]_{3^-}$, and
$[d_{5/2}^{-1}\,d_{5/2}^\Lambda]_{5^+}$ and
$[d_{5/2}^{-1}\,d_{3/2}^\Lambda]_{4^+}$ corresponding to the
proton hole in $d_{5/2}$ and $\Lambda$ in $j_>$ and $j_<$, if
observed experimentally, might give us information on the
spin-orbit splitting of the $\Lambda$ $p$- and $d$-states,
respectively.

The photo-production reactions with $^{40}{\rm Ca}$ and $^{52}{\rm
Cr}$ targets are discussed as other interesting cases. The former
nucleus is $LS$-closed up to $j_{<}=0d_{3/2}$ shell orbit, so that
the highest spin in the $[d_{3/2}^{-1}\,j_{>}^\Lambda]_J$
multiplet is $J^\prime_{\rm max}=j_{<}+j_{>}^\Lambda =\ell_{\rm p}
+\ell_\Lambda= L_{\rm max}$ with a natural parity and such a state
is strongly excited. However, the states of $[d_{3/2}^{-1}
j_{<}^\Lambda]_J\,$-type with $J_{\rm max}^{\prime\prime} =
\ell_{\rm p} + \ell_\Lambda -1 = L_{\rm max} -1$ have weak
strengths in the excitation function. The $j_{<}\,$-closed shell
nuclear target presents the dominant peaks of natural parity,
i.e., $2^+, 3^-$ and $4^+$ as seen in $^{40}_{\,\Lambda}{\rm K}$.
The present case seems to be a nice example of getting the
$\Lambda$ single-particle energies in this mass region which has
not been explored with good energy resolution.

The $^{52}{\rm Cr}$ target contains four protons in an active
$0f_{7/2}$ shell orbit. The dominant peak series in the $(\gamma,
{\rm K}^+)$ reaction consists of unnatural parity states of
$[f_{7/2}^{-1}\,j_{>}^\Lambda]_J\,$-type with $J_{\rm max} =
\ell_{\rm p} + \ell_\Lambda +1 = L_{\rm max} +1$ and $\ell_{\rm p}
= 3$, $\ell_\Lambda = 0, 1$ and 2, respectively, while the
$\Lambda$ spin-orbit partner states
$[f_{7/2}^{-1}\,j_{<}^\Lambda]_{J=L_{\rm max}}$ have reduced
strength as about 60\% of the biggest peak within each multiplet.

Conclusively, the $(\gamma, {\rm K}^+)$ photo-production reactions
on the medium-mass nuclear targets will provide us with novel
and unique features in the $\Lambda$-hypernuclear spectroscopy
through the selective excitation of unnatural parity
highest-spin states (natural parity high-spin states for
the $LS$-closed targets). A possibility is also pointed out
that we might have a chance to investigate $\Lambda$
single-particle energies over whole periodic table including
the spin-orbit splitting for $p$-, $d$-, and $f$-orbits. It is
also notable that the reaction on various nuclear targets produces
proton-deficient hypernuclear species otherwise unaccessible.
Thus the photo- and electro-production of $\Lambda$-hypernuclei
will offer a nice opportunity to get knowledge on the dynamical
behavior of hyperon-nucleus coupling and on the baryon behavior
deeply inside the nucleus.

\begin{acknowledgements}
This work has been done as a part of the Japan-Czech Joint
Research project on ``Electromagnetic Production and Weak Decays
of Hypernuclei'' which was supported for two years by Japan
Society for the Promotion of Science. The work was also supported
in part by Grant-in-aid for Scientific Research ( Specially
Promoted Research 12002001 and Creative Scientific Research
16GS0201 and the 21st Century COE program ``Exploring new science
by bridging particle-matter hierarchy'' by MEXT, Japan ). P.B. and
M.S. acknowledge support by the Grant Agency of the Czech
Republic, grant No. 202/05/2142 and the Institutional Research
Plan AVOZ10480505.
\end{acknowledgements}

%
%  Appendix
%
\appendix*
\section{Expressions of the reduced effective number
        $\rho(fi; M_f)$}
%\vskip 1mm
%\centerline{\bf Expressions of the reduced effective number
%        $\rho(fi; M_f)$ }
%\vskip 1mm

The product of incident $\gamma$ and outgoing K meson distorted
waves is expanded in partial waves
\begin{equation}
 \chi^{(-)*}(\mbox{\bf $p_{\rm K}$},\mbox{\bf $r$})\,
 \chi^{(+)}(\mbox{\bf $p_\gamma$},\mbox{\bf $r$})
 = \sum_{km} \sqrt{4\pi[k]} \ i^k  ~\tilde{j}_{km}(p_\gamma,
   p_{\rm K}, \theta ; \ r) ~Y_{km}(\mbox{\bf $\hat{r}$}) \ .
\end{equation}
Then the $\rho^{gg+}(fi;M_f)$, $\rho^{gg-}(fi;M_f)$ and
$\rho^{gg+-}(fi;M_f)$ are expressed as follows
\begin{equation}
  \rho^{gg+}(fi;M_f) \ \equiv \ \rho^{ggx}(fi;M_f) \quad {\rm with} \
    \ x=1 \ ,
\end{equation}
\begin{equation}
  \rho^{gg-}(fi;M_f) \ \equiv \ \rho^{ggx}(fi;M_f) \quad {\rm with} \
    \ x=-1 \ ,
\end{equation}
and
\begin{eqnarray}
  \rho^{ggx}(fi;M_f) &=& \frac {4\pi}{2[J_i]} \sum_{k_1 k_2}
   \sum_{K_1 K_2} \sqrt{[k_1] [k_2] [K_1] [K_2]}
   ~(-1)^{k_1 +k_2 -K_1 -K_2} ~(-1)^{J_i -M_f} ~i^{k_1 -k_2}
   \nonumber\\
  & \times & \sum_{ \kappa p^\prime p} \
    W(K_1 K_2 J_f J_f ; \kappa J_i) ~(J_f M_f J_f -M_f | \kappa 0)
    ~W(K_1 K_2 k_1 k_2; p^\prime 1)  \nonumber \\
  & \times& \sum_{m} \ (K_1 ~-(m+x) ~K_2\, m+x ~| \kappa 0)
    ~(K_1 m+x ~K_2 ~-(m+x) ~| p^\prime 0)  \nonumber \\
  &  & \qquad \quad  \cdot \ (k_1 -m ~k_2 m | p^\prime 0)
             ~(k_1 m k_2 -m | p 0) \nonumber \\
  & \times & \sum_{m^\prime m^{\prime\prime}}
         (-1)^{m^{\prime\prime}}
         (k_1 m^\prime k_2 -m^{\prime\prime} | p q) \ i^{|q|}
         ~d(p,q)  \nonumber \\
  &  & \qquad  \cdot \  \langle J_f \parallel
        \tilde{j}_{k_1 m^\prime}(p_\gamma, p_{\rm K}, \theta, r)
        ~[Y_{k_1} \times \mbox{\bf $\sigma$} ]_{K_1}
         \parallel J_i \rangle_{S_1}  \nonumber \\
  &  & \qquad \quad \cdot \ \langle J_f \parallel
        \tilde{j}_{k_2 m^{\prime\prime}}(p_\gamma, p_{\rm K}, \theta, r)
        ~[Y_{k_2} \times \mbox{\bf $\sigma$} ]_{K_2}
         \parallel J_i \rangle^{*}_{S_1}\ ,
\end{eqnarray}
where $d(p,q)$ is a function defined in Ref.~\cite{Bando}.
\begin{eqnarray}
  \rho^{gg+-}(fi;M_f) &=& \frac {4\pi}{2[J_i]} \sum_{k_1 k_2}
   \sum_{K_1 K_2} \sqrt{[k_1] [k_2] [K_1] [K_2]}
   ~(-1)^{k_1 -K_1} ~(-1)^{J_i -M_f}\,i^{k_1 -k_2}
   \nonumber\\
  & \times & \sum_{ \kappa p^\prime p} \
    W(K_1 K_2 J_f J_f ; \kappa J_i) ~(J_f M_f J_f -M_f | \kappa 0)
    ~W(K_1 K_2 k_1 k_2; p^\prime 1)  \nonumber \\
  & \times& \sum_{m} \ (K_1 ~-(m+1) ~K_2 m+1 ~| \kappa 0)
    ~(K_1 m+1 ~K_2 m+1) ~| p^\prime 2m +2)  \nonumber \\
  &  & \qquad \quad  \cdot \ (k_1 -m ~k_2 ~-m-2 | p^\prime -2m -2)
             ~(k_1 m k_2 ~-m-2 | p -2) \nonumber \\
  & \times & \sum_{m^\prime m^{\prime\prime}}
         (-1)^{m^{\prime\prime}}
         ~(k_1 m^\prime k_2 -m^{\prime\prime} | p q) \ i^{q}
         ~d^{p}_{-q,2}(\frac {\pi}{2})  \nonumber \\
  &  & \qquad  \cdot \  \langle J_f \parallel
       \tilde{j}_{k_1 m^\prime}(p_\gamma, p_{\rm K}, \theta, r)
        ~[Y_{k_1} \times \mbox{\bf $\sigma$} ]_{K_1}
         \parallel J_i \rangle_{S_1}  \nonumber \\
  &  & \qquad \quad \cdot \ \langle J_f \parallel
        \tilde{j}_{k_2 m^{\prime\prime}}(p_\gamma, p_{\rm K}, \theta, r)
        ~[Y_{k_2} \times \mbox{\bf $\sigma$} ]_{K_2}
         \parallel J_i \rangle^{*}_{S_1} \ ,
\end{eqnarray}
where $d^{j}_{m m^\prime}(\theta)$ is a rotation matrix~\cite{Edmonds}.

%
%  References
%

%
%
%%% Tables
%
%
% Table 1
%
\begin{table}[hbt]
\caption{Comparison of the spin-independent ($f_0$) and spin-dependent
parts ($g_0$, $g_{1}$, $g_{-1}$) of the elementary amplitude is shown
for the four models adopted in this paper. The values are evaluated for
two kaon laboratory angles at $E_{\gamma}^{\rm L}=1.3$ GeV.
Units of $|f_0|^2$, $|g's|^2$, and Re($f_0g_0^*$) are $\mu$b/sr/GeV$^4$.
The laboratory cross sections are in $\mu$b/sr. }
\label{tab31gk}
\smallskip
\begin{tabular}{ccrrrrrcc}
\hline\hline\vspace{1mm}
$\theta^{\rm L}_{\rm K}$ & Model & $|f_0|^2$& $|g_0|^2$& $|g_{1}|^2$
& $|g_{-1}|^2$&Re($f_0g_0^*$)& d$\sigma$/d$\Omega$ & Pol($\Lambda$)\\
\hline
      & AS1 & 0.0067 & 0.374 & 0.179 & 0.193 & $-$0.0141 & 1.91 & $-$0.055\\
3$^\circ$& KMAID & 0.0149 & 0.296 & 0.148 & 0.148 &$-$0.0642 & 1.54 & $-$0.212\\
      & C4  & 0.0002 & 0.572 & 0.272 & 0.281 & $-$0.0063 & 2.85 & $-$0.019\\
      & SLA & 0.0023 & 0.424 & 0.211 & 0.208 & $-$0.0079 & 2.14 & $-$0.016\\
\hline
      & AS1 & 0.0515 & 0.352 & 0.164 & 0.192 & $-$0.0397 & 1.84 & $-$0.142\\
10$^\circ$&KMAID & 0.1205 & 0.267 & 0.121 & 0.129 &$-$0.1793 & 1.55 & $-$0.575\\
      & C4  & 0.0010 & 0.534 & 0.179 & 0.204 & $-$0.0179 & 2.23 & $-$0.066\\
      & SLA & 0.0239 & 0.368 & 0.175 & 0.167 & $-$0.0235 & 1.78 & $-$0.053\\
\hline\hline
\end{tabular}
\end{table}

%
% Table 2
%
%%%%%%%%%( Table 2: New with DDHF values, Jan_2007 )%%%%%%%%%%%%%%%%%%
\newpage
\begin{center}
\squeezetable
\begin{table}
\caption{Differential cross sections (in nb/sr) for the
$^{28}$Si($\gamma,{\rm K}^+)^{28}_{\,\Lambda}$Al reaction
calculated in DWIA (a) at $E^{\rm L}_{\gamma}=1.3$ GeV and
$\theta_{\rm K}^{\rm L}=3^\circ$ with the Saclay-Lyon A
amplitude~\cite{SLA98}. The final hypernuclear states are
expressed by $[(nlj)_{\rm p}^{-1}(nlj)^{\Lambda}]_J$. The
$\Lambda$ DDHF single-particle  energies ($E_{\Lambda}$) in MeV
are listed in the parentheses where the $\Lambda$N spin-orbit
interaction is neglected for simplicity. In the lower section (b),
the estimates without Kaon distortion are listed for comparison in
the case of $[(0d_{5/2}^{-1})_{\rm p}(nlj)^{\Lambda}]_J$. }
\label{Tab:Si28d5}
\begin{tabular}{rrrrrrr}
\hline\hline
(a) DWIA        % & & & & & & \\
 & $0s_{1/2}^{\Lambda}$& $0p_{3/2}^{\Lambda}$ & $0p_{1/2}^{\Lambda}$
 &$0d_{5/2}^{\Lambda}$& $0d_{3/2}^{\Lambda}$ & $1s_{1/2}^{\Lambda}$ \\
 ($E_{\Lambda}$)&($-$16.92)&($-$8.40)&($-$8.40)&(0.69)&(0.69)&(0.32)\\
 \hline
 (p-hole)\, $J$=0& $-$ & $-$ & $-$  &  0.1& $-$ & $-$\\
                   $J$=1& $-$ &  5.3& $-$  & 27.6&  7.2& $-$\\
$(0d_{5/2}^{-1})$ $J$=2& 32.9&  2.9&  15.7&  0.2& 33.6& 1.6\\
                   $J$=3& 65.5&  4.6&  87.0& 20.6& 31.1& 3.1\\
                   $J$=4& $-$ &141.0& $-$  &  0.3&124.1& $-$\\
                   $J$=5& $-$ & $-$ & $-$  &165.4& $-$ & $-$\\
\hline
                   $J$=0& 0.0 & $-$ &  0.0 & $-$ & $-$ & 0.0\\
$(0p_{1/2}^{-1})$ $J$=1&38.6 &  0.6& 27.5 & $-$ &  1.1&19.8\\
                   $J$=2& $-$ & 75.0& $-$  &  8.6& 38.2& $-$\\
                   $J$=3& $-$ & $-$ & $-$  & 87.4& $-$ & $-$\\
\hline
                   $J$=0& $-$ &  0.0& $-$ & $-$ &  0.0& $-$\\
                   $J$=1& 17.8& 11.2&  0.5&  2.7&  4.8&  9.7\\
$(0p_{3/2}^{-1})$ $J$=2& 52.8&  0.2& 68.6&  1.8& 15.7& 29.0\\
                   $J$=3& $-$ &110.6& $-$ &  4.8&108.8& $-$\\
                   $J$=4& $-$ & $-$ & $-$ &145.4& $-$ & $-$ \\
\hline
                   $J$=0& 0.0 & $-$ &  0.0& $-$ & $-$ &  0.0\\
$(0s_{1/2}^{-1})$ $J$=1&18.0 & 14.8& 29.4& $-$ & 14.7& 56.8\\
                   $J$=2& $-$ & 44.0& $-$ & 29.6& 44.1& $-$\\
                   $J$=3& $-$ & $-$  &  $-$& 58.7& $-$ & $-$\\
\hline
(b) PWIA         %& & & & & & \\
 & $0s_{1/2}^{\Lambda}$& $0p_{3/2}^{\Lambda}$ & $0p_{1/2}^{\Lambda}$
 &$0d_{5/2}^{\Lambda}$& $0d_{3/2}^{\Lambda}$ & $1s_{1/2}^{\Lambda}$ \\
 ($E_{\Lambda}$)&($-$16.92)&($-$8.40)&($-$8.40)&(0.69)&(0.69)&(0.32)\\
 \hline
 (p-hole)\, $J$=0& $-$ & $-$ & $-$  &  0.2& $-$ & $-$\\
                   $J$=1& $-$ & 10.9& $-$  & 59.6& 15.3& $-$\\
$(0d_{5/2}^{-1})$ $J$=2& 68.3&  4.03& 33.4&  0.4& 71.9& 1.4\\
                   $J$=3&135.6&  8.3& 155.3& 46.5& 58.9& 2.7\\
                   $J$=4& $-$ &250.5& $-$  &  0.4&186.9& $-$\\
                   $J$=5& $-$ & $-$ & $-$  &248.1& $-$ & $-$\\
\hline\hline
\end{tabular}
\end{table}
\end{center}

%
% Table 3
%
%%%%%( Table 3: SLA vs. KMAID for 28Si(g,K+) with (sd)^n wf)%%%%
\begin{center}
\begin{table}
\caption{ Comparison of excitation cross sections of
$^{28}$Si($\gamma,{\rm K}^+)^{28}_{\,\Lambda}$Al calculated using
SLA and KMAID amplitudes at $E^{\rm L}_\gamma =1.3$ GeV and
$\theta_{\rm K}^{\rm L}= 3^\circ$ and 10$^\circ$. Cross sections
are in units of nb/sr.} \label{Tab:Si28Al}
\smallskip
\begin{tabular}{rcrcrc}
 \hline\hline
 {J$^\pi$} & $E_x$  & \multicolumn{2}{c}{($d\sigma/d\Omega$)} &
 \multicolumn{2}{c}{($d\sigma/d\Omega$)} \\
  & [MeV] & \multicolumn{2}{c}{($\theta_{\rm K}=3^\circ$)}&
\multicolumn{2}{c}{($\theta_{\rm K}=10^\circ$)}\\
  &  &SLA&KMAID&SLA&KMAID\\
\hline
2$^+$ & 0.0 & 19.8 & 14.3 &  8.2 &  7.6 \\
3$^+$ & (0.3) & 39.4 & 28.1 & 15.1 & 12.0 \\
4$^-$ & (8.6) & 84.9 & 60.5 & 44.9 & 36.3 \\
3$^-$ & (8.9) & 52.4 & 36.7 & 28.9 & 21.2 \\
5$^+$ & (17.6) & 99.6 & 70.9 & 69.2 & 56.0 \\
4$^+$ & (18.1) & 74.7 & 52.3 & 53.4 & 39.5 \\
\hline
\end{tabular}
\end{table}
\end{center}

%
%
%%  Figures
%
\newpage
%
% Figure 1
%

%
% Figure 2
%

% Figure 3
%

%
% Figure 4
%

%
% Figure 5
%

%
% Figure 6
%

%
% Figure 7
%

%
% Figure 8
%

%
% Figure 9
%

%
% Figure 10
%

%
% Figure 11
%

%
% Figure 12
%

\end{document}